\documentclass[10pt, conference]{IEEEtran}
\usepackage{amsfonts}
\usepackage{amssymb}
\usepackage{eurosym}
\usepackage{cite}
\usepackage{graphicx}
\usepackage{epstopdf}
\usepackage{amsmath}
\usepackage{tikz,lipsum}
\usepackage{caption}
\usepackage[T1]{fontenc}
\usepackage{amsthm}
\usepackage{mathrsfs}
\usepackage{color}
\usepackage{stfloats}
\usepackage{xcolor, etoolbox}
\usepackage{subcaption}
\usepackage{authblk}

\usepackage{array, multirow, makecell}
\IEEEoverridecommandlockouts

\def\BibTeX{{\rm B\kern-.05em{\sc i\kern-.025em b}\kern-.08em
    T\kern-.1667em\lower.7ex\hbox{E}\kern-.125emX}}
\begin{document}

\title{Analysis of Blockchain Integration in the e-Healthcare Ecosystem}
\author[1]{Abdellah OUAGUID}
\author[2]{Mohamed HANINE}
\author[3]{Zouhair CHIBA}
\author[3]{Noreddine ABGHOUR}
\author[4,5,6]{Hassan GHAZAL}

\affil[1]{ENSET, Hassan II University, Mohammedia, Morocco}
\affil[2]{LTI Laboratory, National School of Applied Sciences, Chouaib Doukkali University, El Jadida, Morocco}
\affil[3]{L.I.S Labs, FSAC, Hassan II University, Casablanca, Morocco}
\affil[4]{Laboratory of Genomics, Bioinformatics and Digital Health,  \protect\\Mohammed VI University of Health Sciences, Casablanca, Morocco}
\affil[5]{Laboratory of Genomics, Bioinformatics and Digital Health,  \protect\\Mohammed VI Center for Research and Innovation, Rabat, Morocco}
\affil[6]{Royal Institute of Sports, Salé, Morocco}

\affil[ ]{Emails: ouaguid@gmail.com, hanine.m@ucd.ac.ma, zouhair.chiba@univh2c.ma,  \protect\\ nabghour@gmail.com,  hassan.ghazal@fulbrightmail.org}
\maketitle

\makeatletter
\def\ps@IEEEtitlepagestyle{	
	\def\@oddfoot{\mycopyrightnotice}	
	\def\@evenfoot{}	
}
\def\mycopyrightnotice{	
	{\footnotesize \textbf{979-8-3503-2939-1/23/\$31.00~\copyright2023 IEEE}\hfill}	
	\gdef\mycopyrightnotice{}	
}

\begin{abstract}
No one can dispute the disruptive impact of blockchain technology, which has long been considered one of the major revolutions of contemporary times. Its integration into the healthcare ecosystem has helped overcome numerous difficulties and constraints faced by healthcare systems. This has been notably demonstrated in the meticulous management of electronic health records (EHR) and their access rights, as well as in its capabilities in terms of security, scalability, flexibility, and interoperability with other systems. This article undertakes the study and analysis of the most commonly adopted approaches in healthcare data management systems using blockchain technology. An evaluation is then conducted based on a set of observed common characteristics, distinguishing one approach from the others. The results of this analysis highlight the advantages and limitations of each approach, thus facilitating the choice of the method best suited to the readers' specific case study. Furthermore, for effective implementation in the context of e-health, we emphasize the existence of crucial challenges, such as the incomplete representation of major stakeholders in the blockchain network, the lack of regulatory flexibility to ensure legal interoperability by country, and the insufficient integration of an official regulatory authority ensuring compliance with ethical and legal standards. To address these challenges, it is necessary to establish close collaboration between regulators, technology developers, and healthcare stakeholders.
\end{abstract}


\section{Introduction}
\begin{figure*}
    \centering
    \includegraphics[width=0.80\linewidth]{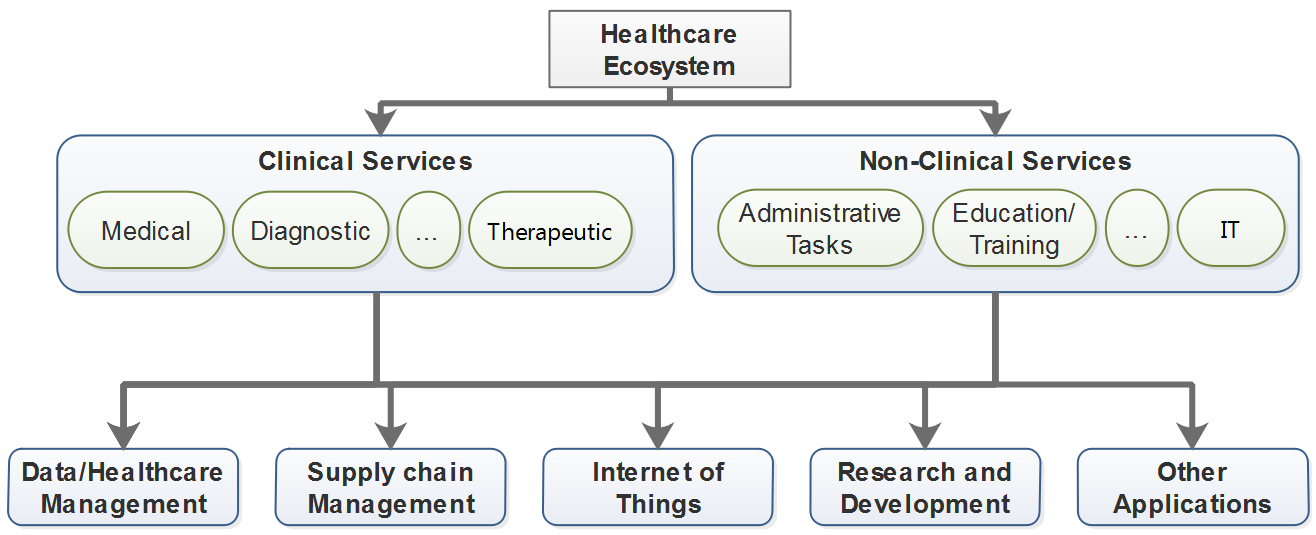}
    \caption{Healthcare Ecosystem Overview}
    \label{fig:healthcare_eco}
\end{figure*}

In the realm of e-health systems, particularly those managing sensitive data such as patients' Electronic Health Records (EHR), play a crucial role in the security and evolution of the healthcare ecosystem\cite{cerchione2023blockchain}. Their continual adaptation to emerging technologies (AI, IoT, blockchain, etc.) enables them to leverage the benefits of these advancements while enhancing their level of maturity.\\
However, the multitude of services comprising the healthcare ecosystem can render the complete digitization of such a system complex, particularly in terms of managing its interoperability with existing or third-party systems. Figure \ref{fig:healthcare_eco} illustrates the organization of services within the healthcare ecosystem, which can be classified into two types: clinical and non-clinical services. Clinical services encompass all activities involving direct interaction with the patient (such as diagnosis, therapy, and observation). In contrast, non-clinical services refer to roles that do not directly involve patient treatment but actively interact within the clinical environment, such as medical research, information technology, and administrative assistance. These diverse services generate data flows, both coherent and incoherent, which fuel the operation of several systems, including EHR management systems, supply chains, and scientific research.\\
The significant transformation experienced in the healthcare sector, particularly during the COVID-19 public health crisis, has attracted the attention of malicious actors, leading to an increase of over 69\% in the volume of ransomware cyber-attacks targeting healthcare systems\cite{TR2021}. This percentage represents the highest proportion compared to all other domains and sectors in 2021. During the first quarter of 2023, industry-specific attack ranking revealed a notable 22\% increase in attacks on the healthcare sector compared to the previous year, with a weekly average of 1,684 attacks (Table \ref{tab:ransomware_attacks})\cite{cpr2023}. This underscores the imperative to prioritize the security, integrity, and confidentiality of patient data while digitizing an electronic healthcare system to mitigate any form of attack (DDoS, data breaches, ransomware, malware, phishing, etc.) that may compromise its proper functioning.
\begin{table}[!t]
\renewcommand{\arraystretch}{1.3}
\caption{Global and Ransomware Attack Rates by Industry (Average Weekly)}
\label{tab:ransomware_attacks}
\centering
\begin{tabular}{|c|c|c|c|c|c|}
\hline
\multirow{2}{*}{Industry} & \multicolumn{4}{c|}{Global Attacks} & \multirow{2}{*}{\makecell{Ransomware\\ Attacks$^*$}} \\
\cline{2-5}
  & \makecell{2022\\Q1} & \makecell{2023\\Q1} & \# & \% &  \\
\hline
Education/Research & 2180 & 2507 & 327 & +15 & 1 out of 26 \\
Government/military & 1675 & 1725 & 50 & +3 & 1 out of 20 \\
Healthcare & 1380 & 1684 & 304 & +22 & 1 out of 27 \\
Communications & 1466 & 1598 & 132 & +9 & 1 out of 33 \\
ISP/MSP & 1474 & 1312 & -162 & -11 & 1 out of 27 \\
Finance/Banking & 1112 & 1212 & 100 & +9 & 1 out of 25 \\
Utilities & 1013 & 1185 & 172 & +17 & 1 out of 32 \\
Retail/Wholesale & 724 & 1079 & 355 & +49 & 1 out of 44 \\
Insurance/Legal & 934 & 1055 & 121 & +13 & 1 out of 40 \\
Leisure/Hospitality & 959 & 997 & 38 & +4 & 1 out of 51 \\
Manufacturing & 982 & 992 & 10 & +1 & 1 out of 41 \\
SI/VAR/Distributor & 917 & 963 & 46 & +5 & 1 out of 31 \\
Consultant & 699 & 881 & 182 & +26 & 1 out of 33 \\
Transportation & 769 & 784 & 15 & +2 & 1 out of 50 \\
Software vendor & 803 & 763 & -40 & -5 & 1 out of 48 \\
Hardware vendor & 398 & 525 & 127 & +32 & 1 out of 49 \\
\hline
\multicolumn{6}{l}{\small $^*$ \textbf{X} attacks out of every \textbf{Y} organizations.}
\end{tabular}
\end{table}
Presently, the deployment of systems or applications requiring substantial storage space and significant computing power has become accessible to all entities, regardless of their size, owing to the growing and continuous adoption of cloud service providers or SaaS partners vying to offer competitively priced services with superior quality \cite{rai2022cloud}.\\
However, various limitations and challenges associated with the client-server paradigm have emerged, particularly concerning the processing, storage, and sharing of sensitive patient data. Furthermore, these systems pose risks regarding the availability, trust, and integrity of stored data. The benefits offered by innovative technology such as blockchain can address certain requirements left unmet by traditional systems\cite{madejczyk2022protection}. Its adoption and use in the healthcare sector ensure system availability (eliminating any single point of failure risk), guarantee patient data traceability, and facilitate their rapid and secure sharing among system entities.\\
In this article, we highlight the common and distinctive features of the four approaches through an in-depth study, comprehensive evaluation, and overall comparison of the described approaches. The rest of the article is structured as follows: section 2 presents the context of EHRs along with an introduction to blockchain technology. Section 3 provides an overview of the integration of EHRs into blockchain-based systems. Subsequently, we present the approaches under study in section 4, followed by a detailed analysis and synthesis in section 5. Finally, section 6 presents the article's conclusion and future research perspectives.

\section{Electronic Health Records (EHRs)}
EHRs are collections of private and sensitive electronic information concerning a patient's health. Data can be gathered from treatments, diagnoses, or analyses performed directly on the patient, with the results/observations recorded by a healthcare professional (physician, nurse, etc.) in a hospital system or automatically retrieved from wearable or portable devices managed by the patients themselves. Secure sharing and access to EHRs must be ensured for all authorized stakeholders (patients, hospitals, laboratories, research centers, insurers, etc... ), while guaranteeing the availability and reliability of stored and exchanged data, reflecting the current health status of the patient as well as their medical history (vaccinations, surgical interventions, etc...).\\
EHRs are sometimes confused with Electronic Medical Records (EMRs) \cite{habib2010ehrs}. As depicted in Table \ref{tab:emr_vs_ehr}, the scope of the EMR is limited as it is not shareable among the various entities within the healthcare ecosystem. This implies that ultimately, the EMR only covers the scope of the creating entity (e.g., the hospital) and is fuelled by data generated through its interventions and care. In contrast, the EHR provides a comprehensive and detailed view of a patient's health information.
\begin{table}[ht]
    \centering
    \caption{Differences Between Electronic Medical Records (EMR) and Electronic Health Records (EHR) in Healthcare Management}
    \begin{tabular}{|l|c|c|}
    \hline
         & \textbf{EMR} & \textbf{EHR}\\
         \hline
        Healthcare Provider(s) & One & Multiple \\
        \hline
        Patient Record & Stored locally & \makecell{Shared between \\healthcare entities}\\
        \hline
        Access for the Patient & View-only & Editable\\
        \hline
        Patient Record History & \multirow{2}{*}{\makecell{Limited to a single\\ provider's office}} & \multirow{2}{*}{\makecell{Accessible across \\various authorized\\provider's offices}}\\
        \cline{1-1}
        Real-time data Accessibility &  & \\
        \cline{1-1}
        Data Analysis &  & \\
        \hline
        Data Interoperability & No & Yes \\
        \hline
    \end{tabular}
    \label{tab:emr_vs_ehr}
\end{table}
\section{Blockchain Technology}
\subsection{Definitions and concepts}
\begin{figure}[ht]
    \centering
    \includegraphics[width=0.75\linewidth]{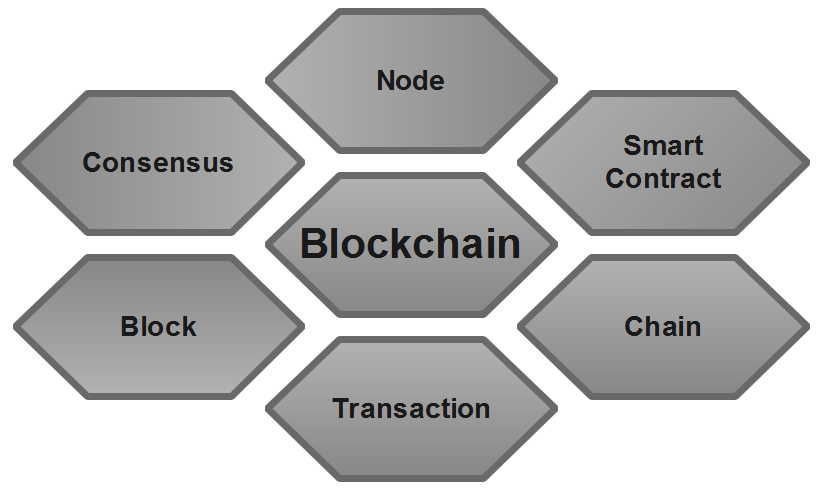}
    \caption{Core components of the Blockchain technology}
    \label{fig:component_bc}
\end{figure}

Blockchain is an emerging secure technology based on a distributed architecture, enabling the secure exchange and storage of data among autonomous '\textit{nodes}', without relying on a central or intermediary authority \cite{padmavathi2023concept}. These nodes, essential constituents of the blockchain network, engage in the exchange, validation, and preservation of data in '\textit{blocks}' that consolidate multiple '\textit{transactions}'. This technology, whose composition is illustrated in Figure \ref{fig:component_bc}, has brought about significant transformations in various sectors, such as finance \cite{ren2023sustainable,zouina2019towards}, the Internet of Things (IoT) \cite{yazdinejad2023secure}, medical applications \cite{al2023performance}, supply chain logistics \cite{hao2023blockchain}, computer security analysis \cite{ouaguid2022androscanreg}, access control systems \cite{amallah2021review}, digital marketing \cite{abirou2022review}, trust management \cite{bellaj2022btrust}, E-learning \cite{bidry2023enhancing} and artificial intelligence \cite{you2022curvetime,el2023unleashing}. Its impact stems from its effective establishment of transparency and trust among participants.\\
The uninterrupted operation of participating nodes (also referred to as '\textit{miners}') ensures the availability, integrity, and reliability of the exchanged and stored data in the '\textit{chain}' by the involved entities, who reach a '\textit{consensus}' on the integrity of these data through consensus algorithms. These algorithms, extensively analyzed in various studies \cite{xiong2022research}, highlight their strengths and limitations. To ensure the stability of a blockchain-based system, each participating node contributes to the system with its resources, including financial stake \cite{39king2012ppcoin} or material resources such as storage \cite{40dziembowski2015proofs}, computing power \cite{41nakamoto2008bitcoin}, etc. The nodes strive to comply with regulations by adhering to pre-established policies and conditions to maintain compliance and avoid exclusion from the network \cite{ouaguid2020node}.\\
Blockchain networks have the capability to host and execute autonomous programs, commonly referred to as '\textit{smart contracts}' \cite{buterin2014next}, which are automatically executed when predefined conditions and terms are met.
\subsection{Data structure}
\begin{figure}
    \centering
    \includegraphics[width=1\linewidth]{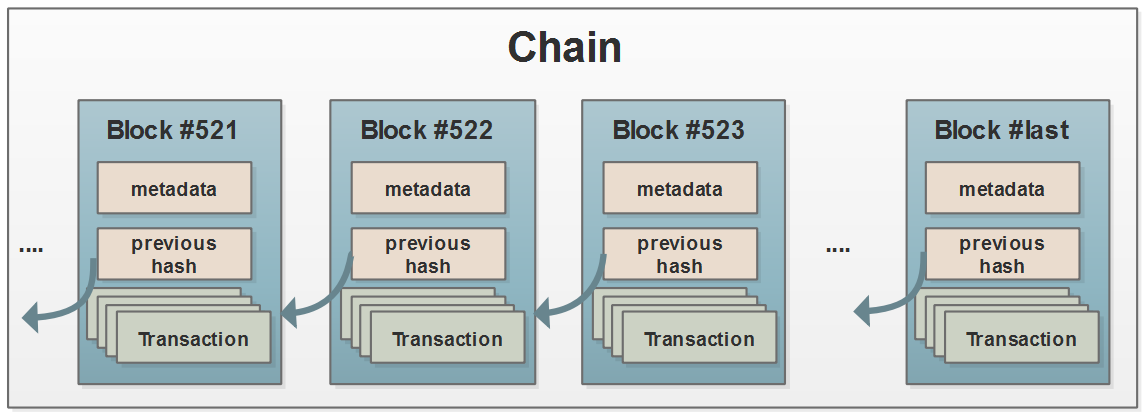}
    \caption{Data structure in Blockchain}
    \label{fig:bc}
\end{figure}

As illustrated in figure \ref{fig:bc}, the design of data stored in the Blockchain is rooted in key principles, namely:
\begin{itemize}
    \item Transaction: This pertains to the operation of adding new data to the chain. Before this is accomplished, the transaction must be exchanged among nodes responsible for grouping them within blocks.
    \item Block: It is a data structure that encapsulates one or more transactions. Generated by system nodes, the block is then broadcast across the network for verification and validation. During each mining round, network nodes record in their local ledgers a single block designated by a consensus algorithm, which is subsequently added to the chain after timestamping and the inclusion of the hash in the valid block preceding it.
    \item Chain: This encompasses all verified and validated blocks since the system's inception. Known for its high availability (hosted on all network nodes) and resistance to any modification, whether legitimate or malicious, due to the existing linkage between blocks through their hashes.
\end{itemize}
\subsection{Classification of Blockchain}
Blockchain technologies can be classified into three categories:
\begin{itemize}
    \item Public Blockchain (Permissionless): Members seeking to integrate into the system by viewing, submitting, or validating transactions can access it without restrictions on their identities or the total number of system members. The security of this system is ensured by consensus algorithms such as Proof of Work (PoW), although their significant consumption of time and energy poses a challenge.
    \item Private Blockchain (Permissioned): Suited for organizations desiring complete control over their processes, data confidentiality, and the identification of participants in their blockchain-based system. The exclusivity of the number of participants allows the deployment of efficient, low-energy consumption consensus algorithms, such as Raft and PBFT. Private blockchains are particularly appealing to financial institutions due to their governance, efficiency in management, and audit of deployed data and processes.
    \item Consortium or Federated Blockchain: Systems based on this type of blockchain promote resource sharing and collaboration in complementary or similar domains. Access to information in this system can be granted to identifiable participants, anonymous profiles, or both. while designating a limited number of identifiable participants playing the role of validators. These validators contribute to increasing system scalability and reducing transaction validation latency.
\end{itemize}

\section{Evolution of EHR Blockchain-based systems}
The design of a confidential, secure, and synchronized sharing environment for the different components of the healthcare ecosystem poses a significant challenge for both government authorities and healthcare providers, whether from a legal or technical perspective. The storage, management, and transfer of sensitive EHR information must be handled transparently, providing patients with full control over their own data, including how it is stored, with which providers, and with whom they wish to share it\cite{whitemorocco}. This must be done in accordance with regulations concerning the handling of sensitive data such as HIPAA (Health Insurance Portability and Accountability Act), GDPR (General Data Protection Regulation), PIPL (Personal Information Protection Law), PCI DSS (Payment Card Industry Data Security Standard), and CCPA (California Consumer Privacy Act).\\
In the literature, several EHR management systems have been proposed. For example, Arshdeep and colleagues \cite{bahga2013cloud} suggested a cloud-based system that provides secure access to integrated data while ensuring data interoperability among different stakeholders to address the issue of the lack of standards in health data exchanges. Other systems, such as the dynamic healthcare system based on the Cloud Computing paradigm\cite{vellela2023integrated}, have also emerged. This system ensures the confidentiality of patient data in a reliable, scalable, and secure environment, based on infrastructure tailored to e-health systems.\\
To overcome the limitations inherent in client-server architectures, particularly in terms of availability (to avoid single points of failure -SPOFs-) and the reliability of stored data (to prevent possible intentional alteration caused by malicious or unintentional attacks, such as data corruption following a bug or database crash), several works have adopted a blockchain-based architecture. This simultaneously addresses the ethical and legal requirements associated with the criticality of the healthcare domain \cite{makinde2023integration}, as well as the technical limitations faced by traditional client-server computerized systems. Alam et al.\cite{alam2023overview} proposed a blockchain-based system for recording previously collected data from IoT devices. These data are integrated for further processing by the various authorized actors within the system. On the other hand, MedBloc (a  Blockchain-Based Secure EHR System for Sharing and Accessing Medical Data) \cite{huang2019medbloc} ensures the confidentiality and interoperability of patient data, granting them access to their medical history and the right to grant or revoke access to their EHR. The BCES solution (a blockchain-based eHealth system for auditable EHRs manipulation in cloud environments), proposed by Huang et al.\cite{huang2021blockchain}, enables the verification of the authenticity of EHR manipulations to ensure transparent, permanent, and verifiable traceability by any internal or external auditing entity. Blockchain technology has also been employed in intelligent healthcare supply chain systems\cite{nanda2023medical}. The proposed approach combines the benefits of the IoT with those of the blockchain to ensure transparent traceability of medical supplies and products, aiming to guarantee the authenticity of products and detect those stemming from counterfeiting.

\section{Healthcare System Management project on Blockchain}
In this section, we present approaches that integrate blockchain technology into their healthcare data management ecosystems, their architectures, and their specificities compared to other existing approaches. The objective is not limited to exploring the design and role of each entity but also to conducting an evaluation based on a set of characteristics (distinctive or common) observed in each approach. We will focus on analyzing four EHR management approaches that have presented the necessary technical specifications for the establishment of their architectures.

\subsection{IoT-based EHR system with Blockchain (BC)}
\begin{figure}[ht]
    \centering
    \includegraphics[width=1\linewidth]{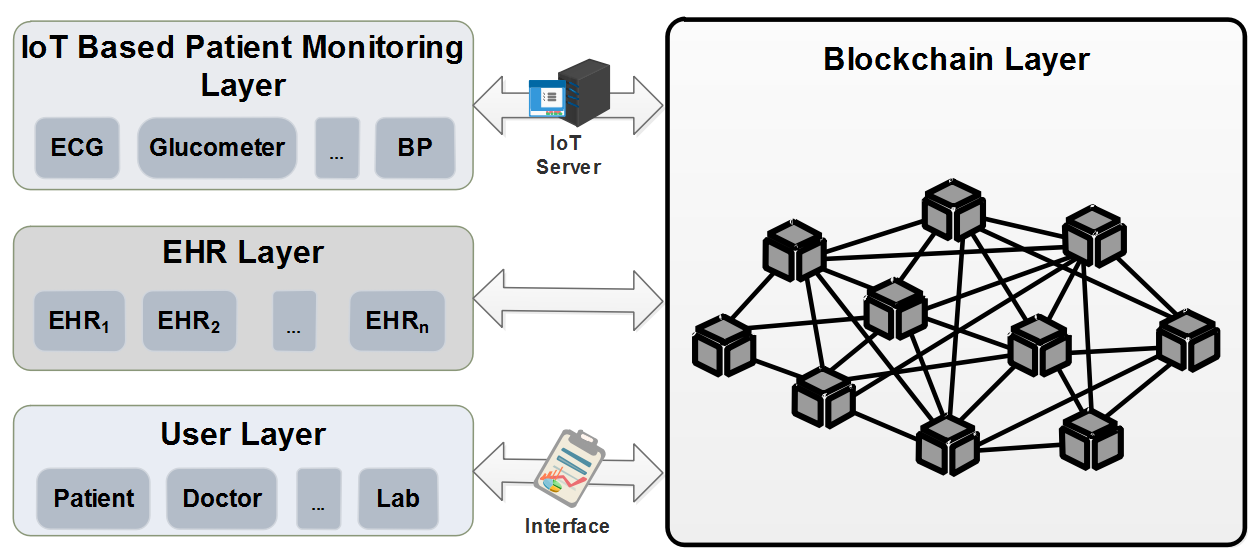}
    \caption{An overview of blockchain-based IoT-EHR framework}
    \label{fig:iotbasedehr}
\end{figure}

Alam et al.\cite{alam2023overview} proposed a blockchain-based framework for IoT-EHR, where patient health data was collected in real-time using IoT devices. The framework also records environmental information, which is integrated with patient data and prepared for future treatment.\\ 
The collected data are stored in a distributed manner, allowing various stakeholders in healthcare provision to seamlessly, securely, and transparently access, visualize, and exchange the patient's EHR data to prescribe appropriate treatments. As illustrated in Figure \ref{fig:iotbasedehr}, the architecture of the proposed approach consists of four layers:

\begin{itemize}
    \item IoT-based patient monitoring layer: The layer of patient sensors comprises a diverse array of sensors aimed at capturing various patient-related data, including but not limited to glucose levels, blood pressure, and body temperature.
    \item EHR layer: At this level, the sharing of specific health records is ensured through collaboration among different healthcare entities and institutions (healthcare organizations, hospitals, medical centers, etc.). The disparity in EHR storage structure and its interoperability is not a challenge, as the data are converted into a unified format via an interface before being communicated to the blockchain layer.
    \item  User layer: Users interact with the system through a dedicated interface that allows them to input or view records in a standardized format, regardless of their initial storage type.
    \item Blockchain (BC) layer: This layer implements several entities managing authentication and verification of records, ensuring the execution of smart contracts and consensus algorithms, guaranteeing the integrity and immutability of local registers safeguarding patients' EHRs, and more.
\end{itemize}
\subsection{MedBloc}
\begin{figure}[ht]
    \centering
    \includegraphics[width=1\linewidth]{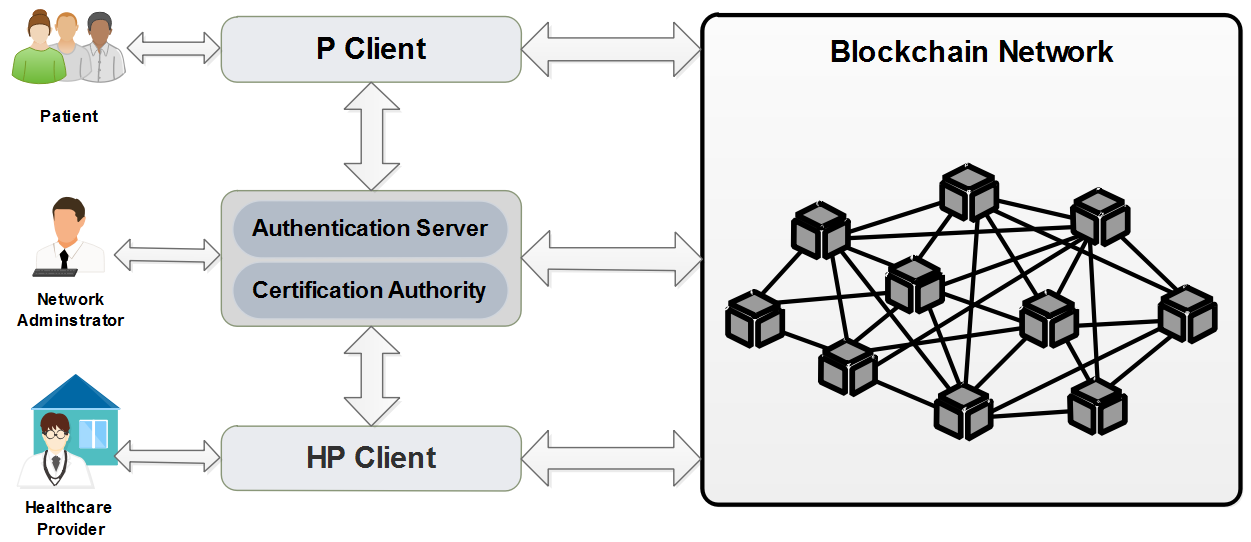}
    \caption{An overview of MedBloc: A blockchain-based secure EHR system}
    \label{fig:medbloc}
\end{figure}

"MedBloc"\cite{huang2019medbloc} refers to a secure EHR system based on a permissioned blockchain developed by Jack Huang and his collaborators. The platform's architecture encompasses various entities within the healthcare ecosystem, including patients (P), certification authorities, healthcare providers (HP), authentication service providers, administrators, and others (Figure \ref{fig:medbloc}).\\
The design of MedBloc aims to improve data interoperability, privacy, and security within the healthcare sector. This translates to transparent sharing and facilitated access to health records for both patients and healthcare providers. The system also allows patients to control access to their medical history by granting or revoking consent to healthcare providers or other authorized entities within the healthcare ecosystem. This approach facilitates informed medical decision-making regarding treatments to be administered.\\
Digital identification and safeguarding of data encryption keys are ensured by entities external to the traditional blockchain. The system utilizes smart contracts to enforce access control rules, thereby preserving patients' confidentiality and privacy.\\
The authors advocated for the use of nontraditional blockchain entities, such as authentication servers and certification authorities. These entities provide the means to issue digital identities and secure encryption keys used to protect data on the blockchain. Finally, smart contracts are deployed to ensure patient confidentiality via access control rules.
\subsection{BCES system}
\begin{figure}[ht]
    \centering
    \includegraphics[width=1\linewidth]{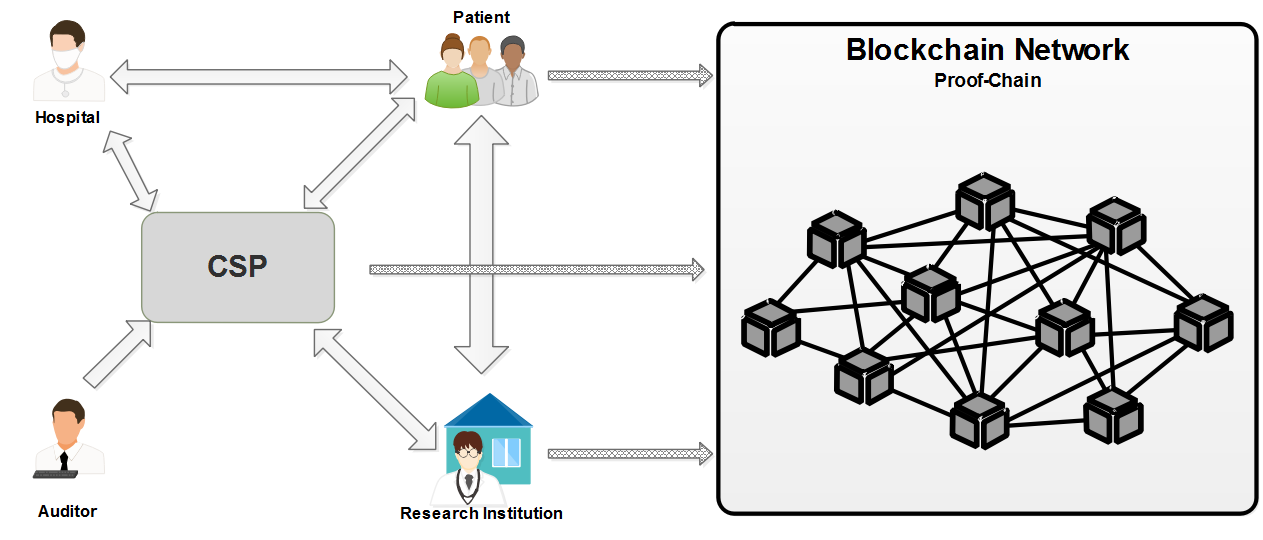}
    \caption{The structure of Blockchain-based eHealth system-BCES}
    \label{fig:BCES}
\end{figure}

In their study, Huang and colleagues\cite{huang2021blockchain} emphasize the importance of preserving data confidentiality during the transmission of EHRs to authorized medical institutions and practitioners. With this goal in mind, they developed the BCES system, a blockchain-based healthcare solution focused on preserving the integrity and availability of EHRs. This system addresses the significant challenge of verifying the authenticity of EHR manipulations. Within the BCES framework, every data modification is meticulously recorded as transactions on the blockchain, ensuring permanent traceability.\\
The proposed architecture (Figure \ref{fig:BCES}) not only ensures data security but also simplifies sharing among care providers by storing metadata on the Blockchain while maintaining large amounts of multimedia data within hospital systems. Key components of BCES include the use of a Proof-Chain mechanism to securely store manipulation logs on the blockchain, enabling transparent tracking of any data alterations. Additionally, the authors introduced an innovative attribute-based proxy re-encryption method for precise control over access to medical data. This cryptographic approach reduces the risks of unauthorized manipulation and ensures data integrity. The implemented encryption techniques combine two methods, attribute-based encryption and proxy re-encryption, strengthening the resilience of the online healthcare system against any unauthorized data modifications, thereby preserving the accuracy of medical diagnoses.\\
In summary, the work of Huang et al. enriches discussions on the applications of blockchain technology in the healthcare domain. They address regulatory challenges and introduce auditing systems tailored to lightweight entities. Their proposal, the BCES system, a blockchain-based healthcare solution, caters to the verification of EHR manipulations. Their innovative approach to blockchain technology, Proof-Chain mechanisms, and attribute-based proxy re-encryption contributes to establishing secure and verifiable EHR management.

\subsection{NAIBHSC}
Blockchain technology has demonstrated its significance in various domains, particularly in pharmaceutical and healthcare supply chains. Nanda et al.\cite{nanda2023medical} introduced a novel approach, known as NAIBHSC, integrating IoT with blockchain in the healthcare supply chain. Its objective is to ensure the tracking of medical product logistics and address issues related to their security, transparency of origin, and real cost, from their source to their final delivery to the consumer. This approach combines the advantages of Blockchain and IoT to create an intelligent healthcare supply chain management system, ensuring easy detection of counterfeit products, which positively impacts the confidentiality, visibility, and trust of various stakeholders in this decentralized system of medical products.\\
\begin{figure}[ht]
    \centering
    \includegraphics[width=1\linewidth]{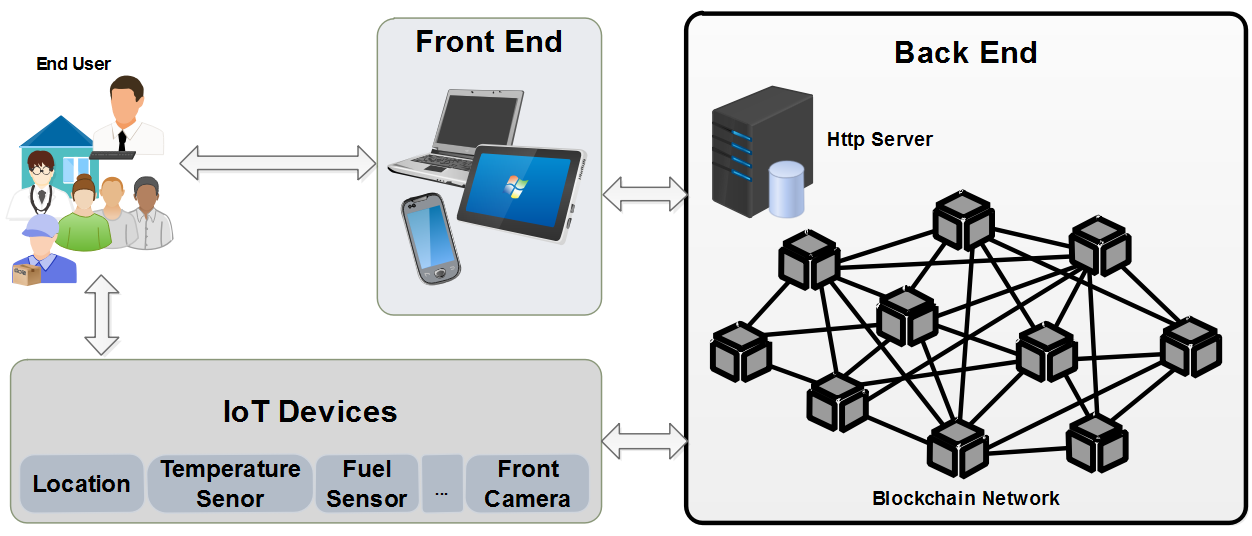}
    \caption{Overview of the NAIBHSC Architecture}
    \label{fig:NAIBHSC}
\end{figure}
As presented in figure \ref{fig:NAIBHSC}, the NAIBHSC architecture comprises the following elements:
\begin{itemize}
    \item IoT Devices: These devices play a crucial role in medical logistics, utilizing sensors to monitor environmental conditions and transmit data to the blockchain network for verification. Before shipment, products are equipped with QR codes and sensors for continuous tracking. Sensors regularly store recorded data, which are later verified by smart contracts. Environmental changes detected during shipment are immediately communicated to relevant stakeholders.
    \item Front End: It comprises diverse user interface devices, such as mobile devices, computers, and others, through which users engage with the back end using a REST API and JSON interface. 
    \item Back End: It is composed of several key components :
    \begin{itemize}
        \item Ethereum Blockchain network: It provides open and user-friendly services, verifying data recorded at the front end, with smart contracts operating within an Ethereum Blockchain virtual machine to ensure data verification and define operations for access control. Users interact with these smart contracts via a REST API and JSON interface, enabling the identification and authentication of user requests, as well as the management of data access permissions.
        \item HTTP Server: processes requests and exchanges information on the World Wide Web (WWW), acting as an interface between the front end and back end, storing, processing, and delivering web pages to clients.
    \end{itemize}
\end{itemize}
Additionally, the authors have highlighted experimental results indicating that the NAIBHSC approach reduces latency, improves response time, and enhances the overall performance of the intelligent healthcare supply chain management system. They have also mentioned the potential of blockchain to revolutionize the structure, planning, management, and operations of the supply chain.

\section{Analysis and Synthesis}
We conducted a systematic comparative analysis of the previously discussed e-health projects based on blockchain. Focusing on their essential key characteristics, we provide a comprehensive evaluation to highlight the specific functionalities of each approach using a detailed summary table (Table \ref{tab:comparative}).
\begin{table*}[htbp]
\caption{A comparative analysis of studied approaches}
\begin{center}
\begin{tabular}{|c|c|c|c|c|}
\hline
& \textbf{BC-based IoT-EHR}& \textbf{MedBloc}& \textbf{BCES} & \textbf{NAIBHSC}\\
\hline
EHR logs traceability/Manipulation traceability & no & no & yes & no \\
\hline
External Audit Entity & no & no & yes & no\\
\hline
Data anonymity & - & yes & - & -\\
\hline
Local register Encrypted & no & yes& - & -\\
\hline
Nodes represent all entities & no & no & no & no\\
\hline
Use Smart Contract & yes & yes & no & yes\\
\hline
Type of storage & on-chain & on-chain & off-chain & on-chain/off-chain \\
\hline
Official regulatory authority & no  & Certification Authority   & Auditor  & no\\
\hline
\end{tabular}
\label{tab:comparative}
\end{center}
\end{table*}

Each of the approaches listed in Table \ref{tab:comparative} is evaluated based on several key criteria:
\begin{itemize}
    \item Traceability of manipulations and EHR logs: BCES is the only approach to offer this functionality, unlike the others that do not provide it.
    \item External Audit Entity: BC-based IoT-EHR, Medbloc, and NAIBHSC do not include an external audit entity. In contrast, BCES ensures that manipulations can be audited by an external entity.
    \item Data Anonymity: Medbloc supports data anonymity, whereas the other approaches do not specify whether they support this feature or not.
    \item Encrypted Local Register: BC-based IoT-EHR stores data encrypted in a local register, whereas for the other approaches, this feature is either not specified or not considered.
    \item Representation of all System Entities: no approach includes all entities of its system as a node in the established blockchain network.
    \item Use of Smart Contracts: Among the proposed systems, only BCES does not employ smart contracts. It is noteworthy that while the described approaches intentionally avoid specifying the programming language for smart contracts, only NAIBHSC explicitly mentions the use of the Solidity language. The authors of these approaches focused on detailing smart contract logic and algorithms without restricting them to a particular language, thereby allowing flexibility in implementation.
    \item Storage Type: BC-based IoT-EHR and Medbloc store data on-chain. The BCES stores data off-chain. However, the storage type adopted by NAIBHSC is hybrid, storing data both on-chain and off-chain.
    \item Official Regulatory Authority as an Entity: BC-based IoT-EHR and NAIBHSC do not include an official regulatory authority as an entity. In contrast, MedBloc specifies the presence of an authority issuing certificates considered identifiers for the participants of the blockchain network, although the official nature of this regulatory entity is not specified. The same observation applies to BCES, which foresees the existence of an auditor allowing the detection of any malicious behavior without specifying the official regulatory authority.    
\end{itemize}

After highlighting the significant differences between the four approaches, we note that for each proposed approach, external actors play primary or secondary roles in the system's operation, without being represented as a node in the blockchain network. Despite the fact that blockchain technology relies on data verification through consensus and shared trust, all system entities (or at least the main entities) cannot participate, as a node, in data validation and therefore contribute to the good governance of the entire ecosystem. This increased dependence on external entities could limit the potential benefits offered by blockchain in a crucial field such as e-health.\\
However, this choice may be motivated by various reasons and challenges that any future approach must consider in the context of e-health. The priority given to a high level of security and confidentiality of electronic patient data, as well as the simplification of the structure and functioning of the blockchain with existing systems, whether internal or external, may have influenced this decision, as highlighted by the Digital Development Agency (ADD) in its proposal \cite{whitemorocco} to use the Blockchain as a secure digital safe for medical data while preserving the current system of each actor involved in patient care, along with its architecture. It is also essential to ensure strong interoperability regarding the technology used and the structure of the exchanged data.\\
Furthermore, the absence of integration of an official regulatory authority within the framework of the studied approaches raises questions. It is worth emphasizing that the integration of an official regulatory authority does not necessarily imply a central authority with absolute power and control. In contrast, a regulatory authority aimed at ensuring adherence to ethical and legal standards could be deployed as an entity represented by at least one node in the network. This authority would have access to various data enabling it to monitor any fraud related to the breach of patient data or falsification of medical records for the purpose of gaining advantage with insurers. This integration also aims to detect any inconsistencies or unlawful collusion among different actors, such as clinics and insurers. Among the practices to be monitored are the unjustified changes in medical service fees, the generation of fictitious invoices, and insurance reimbursements showing inappropriate amounts or frequency. The issue of confidentiality does not arise, as this regulatory authority, potentially governmental or institutional, will have the legal right to access the various information stored in the blockchain. This will be regulated by law or by the explicit consent of the patients themselves.\\
The issue of data interoperability among diverse healthcare platforms and systems is compounded by significant challenges in regulatory flexibility, given the health regulations and jurisdictions associated with the storage, processing, and transfer of sensitive patient data, which vary from one country to another. This situation requires adaptation to the different national regulations of each country as well as international standards.

\section{Conclusion}
The structural and architectural choice of any future e-health system based on blockchain depends on the specific design and objectives of the system. Each requirement may present unique demands in terms of regulatory compliance, security, interoperability, confidentiality, and performance, thereby influencing the implementation and management of the entities within the system. \\
In this article, we specifically examined certain blockchain-based approaches, representing common and individual characteristics related to the context of e-health, including the health supply chain, IoT, and other related areas. \\
The results of our analysis have highlighted the strengths and weaknesses of each approach, emphasizing the importance of various key functionalities in the context of EHR management and the integration of Blockchain technology in this management. However, we have underscored the existence of crucial challenges for effective implementation, necessitating close collaboration among regulators, healthcare stakeholders, and technology developers. These challenges notably include the inadequate representation of key entities in the blockchain network, facilitated by the increasing dependence on external systems, which can compromise effective governance, confidentiality, and data security for patients. We have also emphasized that the appropriate integration of an official regulatory authority is essential to ensure compliance with ethical and legal standards while preserving the decentralized nature of the blockchain. Regulatory flexibility must also be considered to ensure compliance with the jurisdiction of each country using these systems, as well as to guarantee the compliance and interoperability of this system and its data on a global scale.

\bibliographystyle{IEEEtran}
\bibliography{references}

\begin{thebibliography}{10}
\providecommand{\url}[1]{#1}
\csname url@samestyle\endcsname
\providecommand{\newblock}{\relax}
\providecommand{\bibinfo}[2]{#2}
\providecommand{\BIBentrySTDinterwordspacing}{\spaceskip=0pt\relax}
\providecommand{\BIBentryALTinterwordstretchfactor}{4}
\providecommand{\BIBentryALTinterwordspacing}{\spaceskip=\fontdimen2\font plus
\BIBentryALTinterwordstretchfactor\fontdimen3\font minus \fontdimen4\font\relax}
\providecommand{\BIBforeignlanguage}[2]{{%
\expandafter\ifx\csname l@#1\endcsname\relax
\typeout{** WARNING: IEEEtran.bst: No hyphenation pattern has been}%
\typeout{** loaded for the language `#1'. Using the pattern for}%
\typeout{** the default language instead.}%
\else
\language=\csname l@#1\endcsname
\fi
#2}}
\providecommand{\BIBdecl}{\relax}
\BIBdecl

\bibitem{cerchione2023blockchain}
R.~Cerchione, P.~Centobelli, E.~Riccio, S.~Abbate, and E.~Oropallo, ``Blockchain’s coming to hospital to digitalize healthcare services: Designing a distributed electronic health record ecosystem,'' \emph{Technovation}, vol. 120, p. 102480, 2023.

\bibitem{TR2021}
{Thomson Reuters}, ``Ransomware attacks against healthcare organizations nearly doubled in 2021, report says,'' \url{https://www.thomsonreuters.com/en-us/posts/investigation-fraud-and-risk/ransomware-attacks-against-healthcare}, 2022, visited 2023-10-25.

\bibitem{cpr2023}
{Check Point Research}, ``Global cyberattacks continue to rise with africa and apac suffering most,'' \url{https://blog.checkpoint.com/research/global-cyberattacks-continue-to-rise/}, 2023, visited 2023-10-25.

\bibitem{rai2022cloud}
V.~Rai, K.~Bagoria, K.~Mehta, V.~M. Sood, K.~Gupta, L.~Sharma, and M.~Chauhan, ``Cloud computing in healthcare industries: Opportunities and challenges,'' \emph{Recent Innovations in Computing: Proceedings of ICRIC 2021, Volume 2}, pp. 695--707, 2022.

\bibitem{madejczyk2022protection}
M.~Madejczyk, T.~Trejderowski, and B.~Trejderowska, ``Protection of personal data using blockchain,'' \emph{Zeszyty Naukowe Wy{\.z}szej Szko{\l}y Technicznej w Katowicach}, vol.~14, 2022.

\bibitem{habib2010ehrs}
J.~L. Habib, ``Ehrs, meaningful use, and a model emr,'' 2010.

\bibitem{padmavathi2023concept}
U.~Padmavathi and N.~Rajagopalan, ``Concept of blockchain technology and its emergence,'' in \emph{Research Anthology on Convergence of Blockchain, Internet of Things, and Security}.\hskip 1em plus 0.5em minus 0.4em\relax IGI global, 2023, pp. 21--36.

\bibitem{ren2023sustainable}
Y.-S. Ren, C.-Q. Ma, X.-Q. Chen, Y.-T. Lei, and Y.-R. Wang, ``Sustainable finance and blockchain: A systematic review and research agenda,'' \emph{Research in International Business and Finance}, p. 101871, 2023.

\bibitem{zouina2019towards}
M.~Zouina and B.~Outtai, ``Towards a distributed token based payment system using blockchain technology,'' in \emph{2019 International Conference on Advanced Communication Technologies and Networking (CommNet)}.\hskip 1em plus 0.5em minus 0.4em\relax IEEE, 2019, pp. 1--10.

\bibitem{yazdinejad2023secure}
A.~Yazdinejad, A.~Dehghantanha, R.~M. Parizi, G.~Srivastava, and H.~Karimipour, ``Secure intelligent fuzzy blockchain framework: Effective threat detection in iot networks,'' \emph{Computers in Industry}, vol. 144, p. 103801, 2023.

\bibitem{al2023performance}
G.~Al-Sumaidaee, R.~Alkhudary, Z.~Zilic, and A.~Swidan, ``Performance analysis of a private blockchain network built on hyperledger fabric for healthcare,'' \emph{Information Processing \& Management}, vol.~60, no.~2, p. 103160, 2023.

\bibitem{hao2023blockchain}
Y.~Hao, P.~Helo, N.~Tsoniotis, and R.~Toshev, ``Blockchain-based supply chain system in automotive industry forsmall-and medium-sized manufacturing,'' in \emph{Blockchain Driven Supply Chains and Enterprise Information Systems}.\hskip 1em plus 0.5em minus 0.4em\relax Springer, 2023, pp. 151--171.

\bibitem{ouaguid2022androscanreg}
A.~Ouaguid, F.~Fathi, M.~Zouina, M.~Ouzzif, and N.~Abghour, ``Androscanreg 2.0: Enhancement of android applications analysis in a flexible blockchain environment,'' \emph{International Journal of Software Innovation (IJSI)}, vol.~10, no.~1, pp. 1--28, 2022.

\bibitem{amallah2021review}
M.~A. Amallah, N.~Abghour, K.~Moussaid, A.~El~Omri, and M.~Rida, ``Review on blockchain and access control systems,'' in \emph{Advances on Smart and Soft Computing: Proceedings of ICACIn 2021}.\hskip 1em plus 0.5em minus 0.4em\relax Springer, 2021, pp. 235--246.

\bibitem{abirou2022review}
M.~Abirou and N.~Abghour, ``A review of blockchain and the benefits for digital marketing-related applications of blockchain integration,'' \emph{Advances on Smart and Soft Computing: Proceedings of ICACIn 2021}, pp. 355--365, 2022.

\bibitem{bellaj2022btrust}
B.~Bellaj, A.~Ouaddah, E.~Bertin, N.~Crespi, A.~Mezrioui, and K.~Bellaj, ``Btrust: A new blockchain-based trust management protocol for resource sharing,'' \emph{Journal of Network and Systems Management}, vol.~30, no.~4, p.~64, 2022.

\bibitem{bidry2023enhancing}
M.~Bidry, A.~Ouaguid, and M.~Hanine, ``Enhancing e-learning with blockchain: Characteristics, projects, and emerging trends,'' \emph{Future Internet}, vol.~15, no.~9, p. 293, 2023.

\bibitem{you2022curvetime}
J.~You, ``Curvetime: A blockchain framework for artificial intelligence computation,'' \emph{Software Impacts}, vol.~13, p. 100314, 2022.

\bibitem{el2023unleashing}
N.~El~Akrami, M.~Hanine, E.~S. Flores, D.~G. Aray, and I.~Ashraf, ``Unleashing the potential of blockchain and machine learning: Insights and emerging trends from bibliometric analysis,'' \emph{IEEE Access}, 2023.

\bibitem{xiong2022research}
H.~Xiong, M.~Chen, C.~Wu, Y.~Zhao, and W.~Yi, ``Research on progress of blockchain consensus algorithm: A review on recent progress of blockchain consensus algorithms,'' \emph{Future Internet}, vol.~14, no.~2, p.~47, 2022.

\bibitem{39king2012ppcoin}
S.~King and S.~Nadal, ``Ppcoin: Peer-to-peer crypto-currency with proof-of-stake,'' \emph{self-published paper, August}, vol.~19, p.~1, 2012.

\bibitem{40dziembowski2015proofs}
S.~Dziembowski, S.~Faust, V.~Kolmogorov, and K.~Pietrzak, ``Proofs of space,'' in \emph{Annual Cryptology Conference}.\hskip 1em plus 0.5em minus 0.4em\relax Springer, 2015, pp. 585--605.

\bibitem{41nakamoto2008bitcoin}
S.~Nakamoto, ``Bitcoin: A peer-to-peer electronic cash system,'' \emph{Decentralized business review}, 2008.

\bibitem{ouaguid2020node}
A.~Ouaguid, N.~Abghour, and M.~Ouzzif, ``Node security metric: Proof of conformity blockchain consensus protocol,'' in \emph{International Conference on Advanced Intelligent Systems for Sustainable Development}.\hskip 1em plus 0.5em minus 0.4em\relax Springer, 2020, pp. 665--682.

\bibitem{buterin2014next}
V.~Buterin \emph{et~al.}, ``A next-generation smart contract and decentralized application platform,'' \emph{white paper}, vol.~3, no.~37, pp. 2--1, 2014.

\bibitem{whitemorocco}
D.~Anass, E.~A. Hicham, Y.~Azeddine, C.~Saad, and A.~Smail, ``White paper on e-health in morocco: Realities, challenges and development levers,'' \emph{white paper}, 2022.

\bibitem{bahga2013cloud}
A.~Bahga and V.~K. Madisetti, ``A cloud-based approach for interoperable electronic health records (ehrs),'' \emph{IEEE Journal of Biomedical and Health Informatics}, vol.~17, no.~5, pp. 894--906, 2013.

\bibitem{vellela2023integrated}
S.~S. Vellela, B.~V. Reddy, K.~K. Chaitanya, and M.~V. Rao, ``An integrated approach to improve e-healthcare system using dynamic cloud computing platform,'' in \emph{2023 5th International Conference on Smart Systems and Inventive Technology (ICSSIT)}.\hskip 1em plus 0.5em minus 0.4em\relax IEEE, 2023, pp. 776--782.

\bibitem{makinde2023integration}
A.~S. Makinde, A.~O. Agbeyangi, and S.~Omaji, ``Integration of blockchain into medical data security: Key features, use cases, technical challenges, and future directions,'' in \emph{Contemporary Applications of Data Fusion for Advanced Healthcare Informatics}.\hskip 1em plus 0.5em minus 0.4em\relax IGI Global, 2023, pp. 137--165.

\bibitem{alam2023overview}
S.~Alam, S.~Bhatia, M.~Shuaib, M.~M. Khubrani, F.~Alfayez, A.~A. Malibari, and S.~Ahmad, ``An overview of blockchain and iot integration for secure and reliable health records monitoring,'' \emph{Sustainability}, vol.~15, no.~7, p. 5660, 2023.

\bibitem{huang2019medbloc}
J.~Huang, Y.~W. Qi, M.~R. Asghar, A.~Meads, and Y.-C. Tu, ``Medbloc: A blockchain-based secure ehr system for sharing and accessing medical data,'' in \emph{2019 18th IEEE International Conference On Trust, Security And Privacy In Computing And Communications/13th IEEE International Conference On Big Data Science And Engineering (TrustCom/BigDataSE)}.\hskip 1em plus 0.5em minus 0.4em\relax IEEE, 2019, pp. 594--601.

\bibitem{huang2021blockchain}
H.~Huang, X.~Sun, F.~Xiao, P.~Zhu, and W.~Wang, ``Blockchain-based ehealth system for auditable ehrs manipulation in cloud environments,'' \emph{Journal of Parallel and Distributed Computing}, vol. 148, pp. 46--57, 2021.

\bibitem{nanda2023medical}
S.~K. Nanda, S.~K. Panda, and M.~Dash, ``Medical supply chain integrated with blockchain and iot to track the logistics of medical products,'' \emph{Multimedia Tools and Applications}, pp. 1--23, 2023.

\end{thebibliography}

\end{document}